\documentclass[12pt,a4paper]{article}
\usepackage{epsfig}
\usepackage{graphicx}                  
\usepackage{siunitx}
\usepackage{ae}
\usepackage{amsmath}
\usepackage{amssymb}
\usepackage{graphics}
\usepackage{dcolumn}
\usepackage{bm}
\usepackage[symbol]{footmisc}
\begin{document}
\begin{center}

{\Large
{\bf  Stability study and time resolution measurement of Straw Tube detectors}}
\vskip 0.8 true cm
\renewcommand{\thefootnote}{\fnsymbol{footnote}}

{\bf S. Roy$^1 \footnote[1]{Corresponding author} \let\thefootnote\relax\footnote{e-mail: shreyaroy@jcbose.ac.in}$,
S.~Jaiswal$^1 \footnote[2]{Now at Indian Institute of Technology,
Kanpur, Uttar Pradesh 208016, India}$,
S.~Chatterjee$^1$, A.~Sen$^1$, S.~Das$^1$, S.~K.~Ghosh$^1$, S.~Raha$^1$, V.~M.~Lysan$^2$, G.~D.~Kekelidze$^2$, V.~V.~Myalkovsky$^2$, and S.~Biswas$^1$}

\vskip 0.5 true cm

$^{1}${\it Department of Physics and Centre for Astroparticle Physics and Space Science, Bose Institute, EN-80, Sector V, Kolkata-700091, INDIA}

$^{2}${\it LHEP-JINR, Dubna, Russian Federation}
\vskip 0.3 true cm

\end{center}

\abstract{
Straw tube detectors are single wire proportional counters that are widely used as a tracking device. We have carried out R$\&$D with a straw tube detector prototype. The motivation of this work is to study the stability of the performance in terms of gain and energy resolution of the straw tube detectors under high rate radiation. Two different methods are incorporated to perform this study. The gain and energy resolution of the detector are studied along with its variation with ambient temperature and pressure. X-ray from a radioactive source is used to irradiate the detector and the same source is also used to monitor the energy spectra simultaneously for calculation of gain. Variation of the gain and energy resolution of the straw tube detector under X-ray irradiation in Ar/CO$_2$ gas mixture is discussed in this article. We have also estimated the time resolution of the straw tube detectors that can be best achieved with cosmic rays as trigger for the same gas mixture. The details of the measurement process and the experimental results are presented in this article.}

\vskip 0.2 true cm

\begin{center}

Keywords: Straw Tube Detector; Gain; Energy Resolution; Radiation effect; Time Resolution

\end{center}

\section{Introduction}

Straw tube detectors are single wire cylindrical proportional chambers that have been used, in many high energy
physics experiments over the decades such as ATLAS~\cite{atlas}, NA62~\cite{cern} experiment at CERN and GlueX~\cite{glue} experiment in Hall D at JLab for tracking of charged particles
with low material budget. Straw proportional tubes have potential to be used as tracking devices in the future high energy physics experiments~\cite{cbm,fair,vpes} involving very high particle density and extremely high interaction rate. Therefore, it is crucial to test the rate handling capability \cite{sroy1} and the effect of prolonged radiation on the detectors. Earlier studies of straw tubes with Xe/CF$_4$/CO$_2$ gas mixture has revealed gain degradation \cite{mcar,bond}. This transient aging phenomenon observed in strongly irradiated straw tubes is because of change in the gas composition due to a production of long-lived and highly electronegative radicals during the avalanche process. This causes a temporary reduction in gain, which can be restored by an appropriate increase of the gas flow rate. There are more reports on radiation hardness and aging effects of straw tube detectors \cite{tak,kad} performed with Xe based gas mixtures. However, Ar/CO$_2$ is by far a much more widely used mixture in gaseous particle detectors. As far as CO$_2$ gas is concerned, it is believed to be an aging resistant gas unlike other organic gases that are mixed with noble gas to quench secondary photons. Pure Ar/CO$_2$ gas showed stable operation up to $\sim$~1~C/cm \cite{kad,age0,age1,age2}, while some reports showed unexplained gain reduction with this gas~\cite{age3,age4}. Our goal was to operate the straw tube detector under conditions as close as possible to the real environment of high energy physics experiments in terms of total charge accumulated on the detector over its lifetime. We wanted to study the effect of long term exposure to radiation of straw tube detectors and to verify if gas aging phenomena takes place or not without imposing accelerated aging. In this regard, two separate experiments were performed. In the first experiment, stability tests for the absolute gain and energy resolution of one straw tube under X-ray radiation was carried out. The experimental setup and results are discussed in Section~\ref{expsetup} and Section~\ref{res} respectively. The influence of temperature and pressure on the gas gain is also studied. Reduction in gain after continuous operation for a very long time is observed in this experiment. To confirm that this gain reduction is due to the aging phenomena, another experiment using two straw tubes, one as a reference and the other as a test detector was carried out. The reference straw was under the influence of a low rate of X-ray radiation whereas the other one was under a higher rate of X-ray radiation. A comparison of the gain of both the straws was done at certain time intervals during continuous radiation exposure. A detailed description of the experimental setup and discussion on the results are included in Section~\ref{expsetup2} and Section~\ref{res2} respectively. The difference between the first and the second experiment was that, the latter was conducted at relatively lower gas flow rates for reasons discussed in Section~\ref{dis}. Time resolution of the detector is another important factor of concern in any tracking system. We have used cosmic rays as the radiation source and measured the time resolution of the straw tube detector. The experimental setup and results are discussed in Section~\ref{timemeas}.

\section{Stability test of straws : Experiment I}
In this experiment, our main focus is to measure the gain of the straw tube detector continuously at finite intervals and observe its variation with exposure to radiation. Since we know that the gain of a gaseous detector has a dependency on ambient temperature and pressure \cite{altunbus}, we also tried to check their correlation. 
\subsection{Experimental set up}\label{expsetup}
The straw tube prototype used in this experiment is built in JINR, Dubna,
Russia. It consist of 6 straws of diameter 6 mm and length 25~cm.
The straw tube was fabricated from a kapton film, one side containing a conductive layer of
1000-3000 $\si{\angstrom}$ Al + 4 $\mu$m carbon-loaded kapton and the other side containing a thermoplastic 
polyurethane layer of 3 $\mu$m. Two kapton film tapes (4-8~mm wide) were wound in spiral at a 
temperature of around 200~$^{\circ}$C. The thickness of the straw wall is around 65~$\mu$m and the diameter of gold plated tungsten central wire is 30~$\mu$m.
There is a provision to collect signals from each straw through LEMO connector. A premixed gas of Argon and CO$_{2}$ in 80/20 volume
ratio is used in flow mode at a rate of 3 l/hr.

\begin{figure}[h!]
\begin{center}
\includegraphics[scale=0.6]{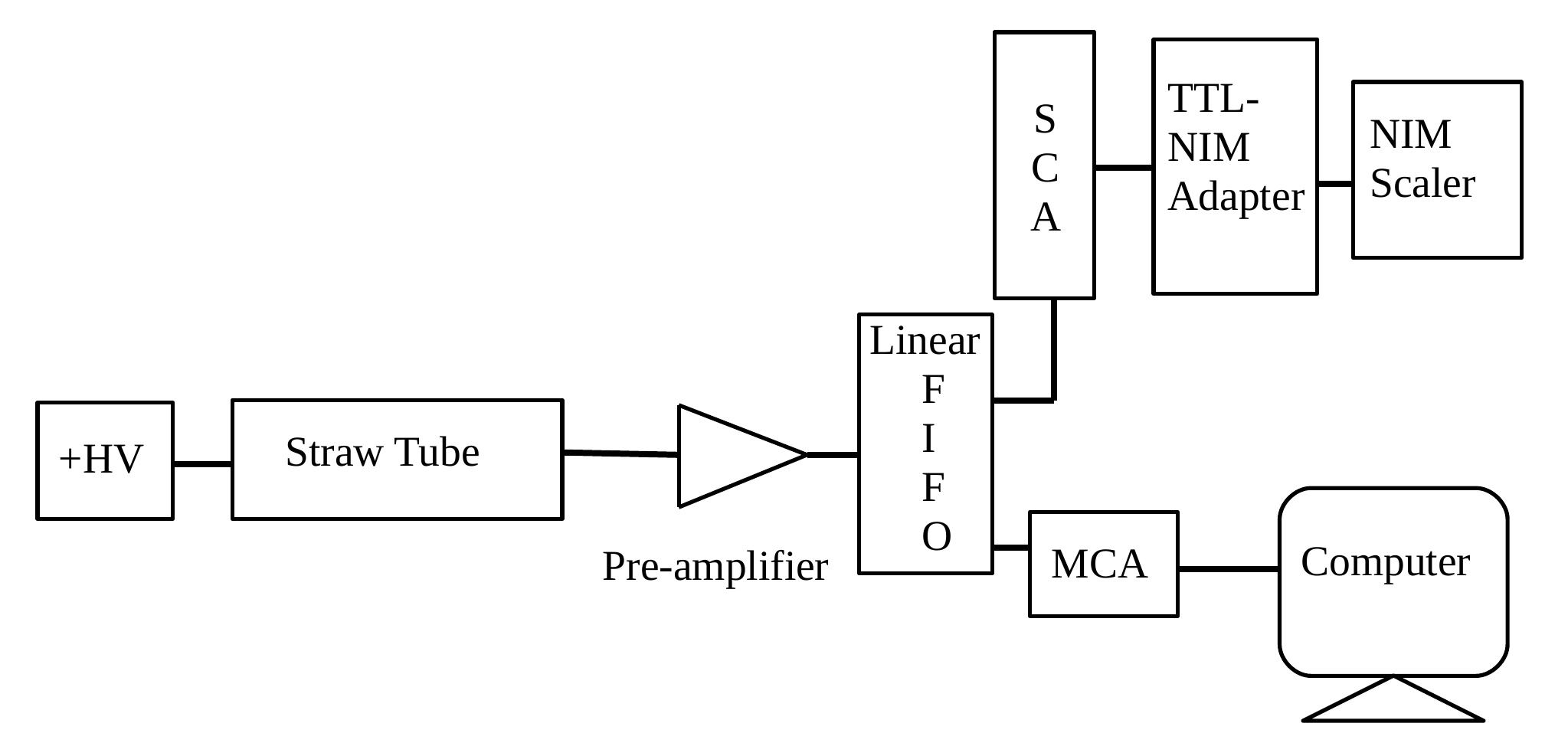}
\caption{Schematic of the setup for the stability test of the straw tube detector.}
\label{strawcircuit}
 \end{center}
\end{figure}

\begin{figure}[h!]
\begin{center}
\includegraphics[scale=0.35]{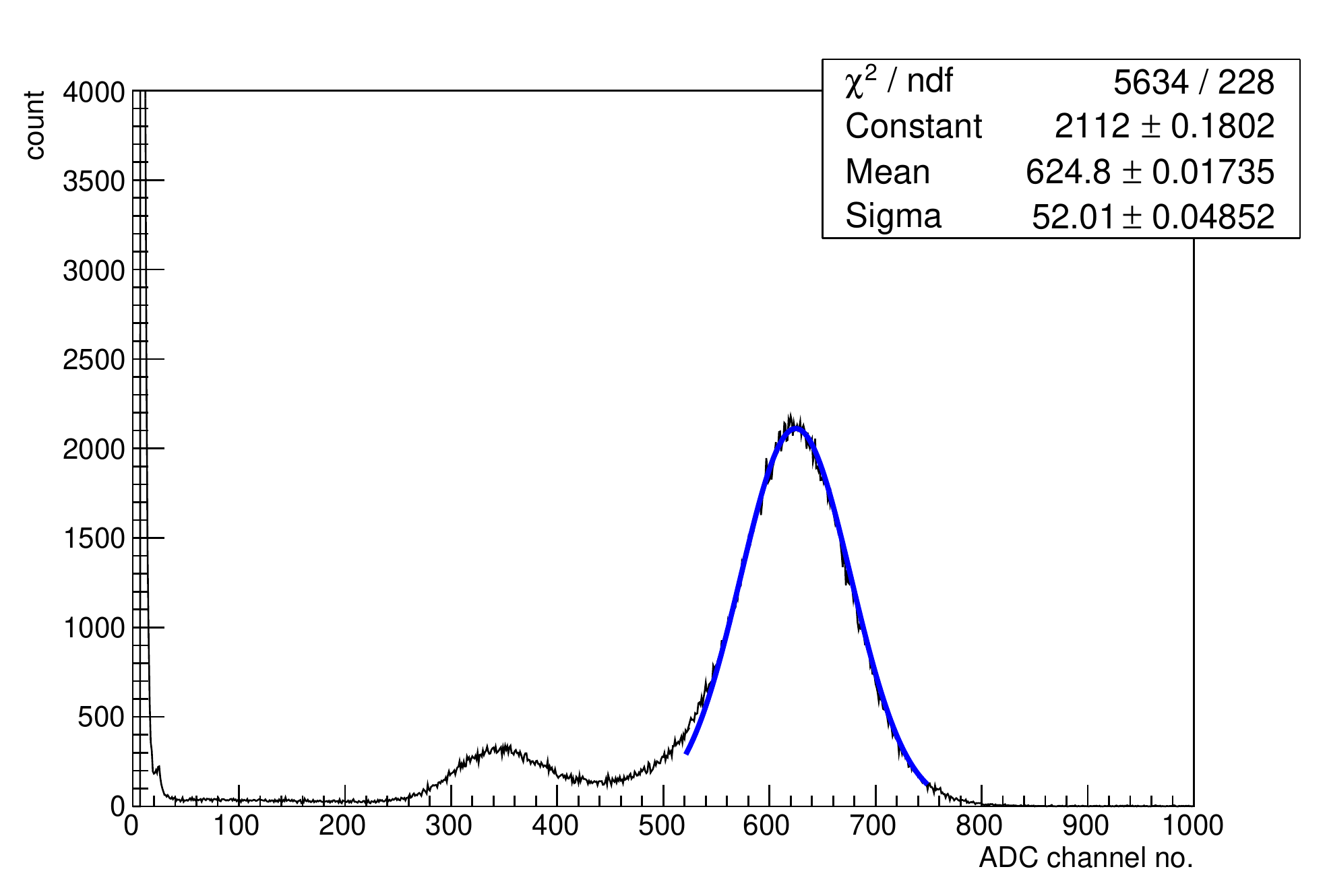}
\caption{Typical energy spectrum for X-ray from Fe$^{55}$ source in Ar/CO$_{2}$ gas mixture of 80/20 volume ratio at 1550~V. The main peak is fitted by a gaussian function shown in blue line.}
\label{spectrum}
 \end{center}
\end{figure}

The central anode wire of the straws are biased with positive high voltage (HV) 
using a HV filter box at one end and the signal is collected from the other end after a capacitor.
The output signal
from the straw is fed to a charge sensitive pre-amplifier having gain 2~mV/fC and shaping time of 300~ns. The output of the pre-amplifier is put to a linear Fan-in-Fan-Out (FIFO) module. One output of the linear FIFO is put to a timing SCA (Single Channel Analyzer), which is operated in integral mode and the
lower level in the SCA is used as the threshold.
A NIM based scalar module is used to measure the count rate from the detector. A Multi Channel Analyser (MCA) is used to obtain the energy spectra with Fe$^{55}$ X-ray source taking another output from the linear FIFO. A schematic of the setup is shown in Fig.~\ref{strawcircuit}. A typical energy spectra for Fe$^{55}$ in Ar/CO$_{2}$ 80/20 mixture at 1550~V is shown in Fig.~\ref{spectrum}.

The gain of the straw tube detector is calculated in the following way. The 5.9~keV peak of the Fe$^{55}$ X-ray spectrum is fitted with a Gaussian function and from the mean of the fitted peak, the charge after avalanche multiplication (output charge) is calculated using the gain of the pre-amplifier (in mV/fC) and the calibration factor ($C. F.$) of the MCA channel number and pulse height (in mV). The expression for gain is given by the ratio of output charge and input charge :
\begin{eqnarray}
 gain &=& \frac{output~charge}{input~charge} \\
 &=& \frac{(Mean \times C. F.)/2 mV)~fC}{No.~of~primary~electrons~\times~e~C} 
 \label{gaineq}
\end{eqnarray} 
The preamplifier used in the set-up offers a gain of 2 mV/fC. The average number of primary electrons produced in the gas is taken as 217 for Ar/CO$_{2}$ 80/20 mixture assuming full energy deposition of 5.9~keV X-ray in the gas volume.
The energy resolution of the straw tube detector is defined as :
\begin{equation}
 \%~energy~resolution = \frac{Sigma~\times~2.355}{Mean} \times 100 \: \%  
\end{equation}
where Sigma and Mean are obtained from the Gaussian fitting of the 5.9~keV peak of each Fe$^{55}$ X-ray spectrum. 
The typical gain of the straw tube detector is found to be ~1.4~$\times$~10$^4$ at a biasing voltage of 1550~V. \\In order to study the effect of prolonged radiation on the detector, a collimated X-ray source is placed on top of the detector and continuous monitoring of the energy spectra with the same source is carried out.
A realistic particle rate of 40 kHz/mm was set using the collimator. The spectra are stored automatically at regular intervals of ten minutes. A data
logger~\cite{datalogger} made in house was used to record ambient temperature and pressure online.
\begin{figure}[h!]
\begin{center}
\includegraphics[scale=0.35]{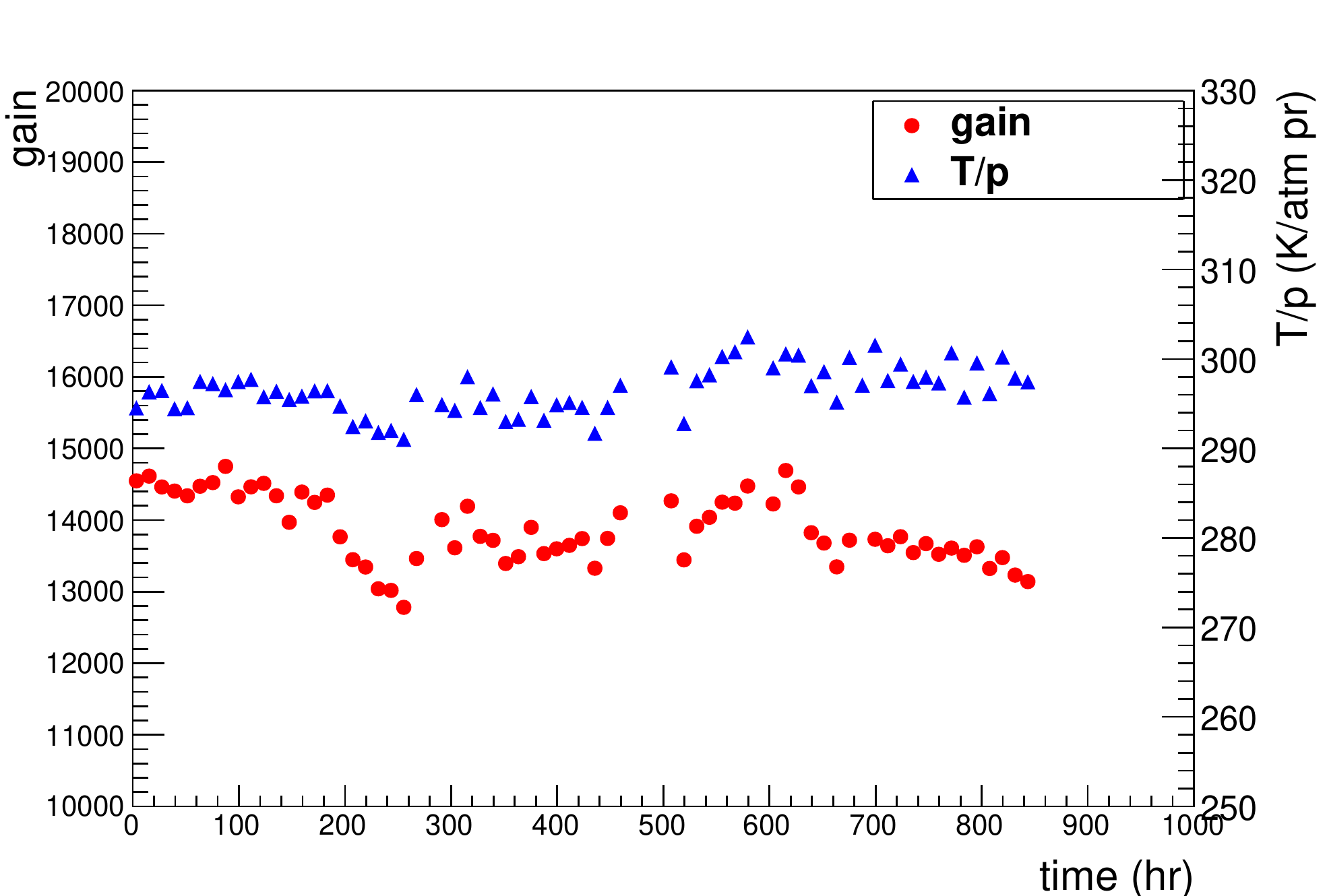}
\caption{Gain and T/p as a function of time. The bias voltage of the straw tube detector is 1550~V.}
\label{gain}
 \end{center}
\end{figure}

\begin{figure}[h!]
\begin{center}
\includegraphics[scale=0.36]{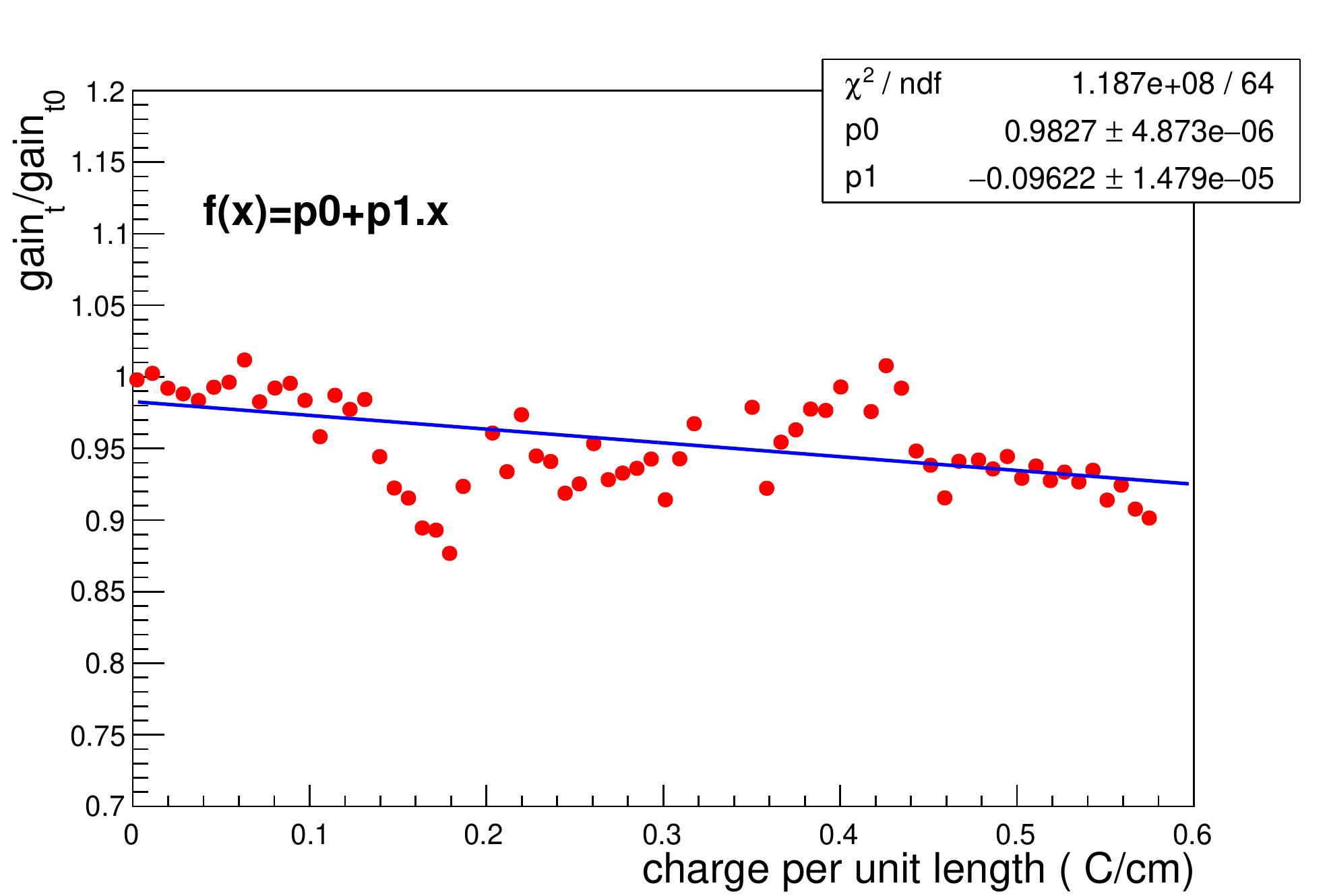}
\caption{Ratio of instantaneous gain and initial gain (normalised gain) as a function of charge accumulated per unit length.}
\label{normgain}
 \end{center}
\end{figure}

\subsection{Results}\label{res}
\begin{figure}[h!]
\begin{center}
\includegraphics[scale=0.35]{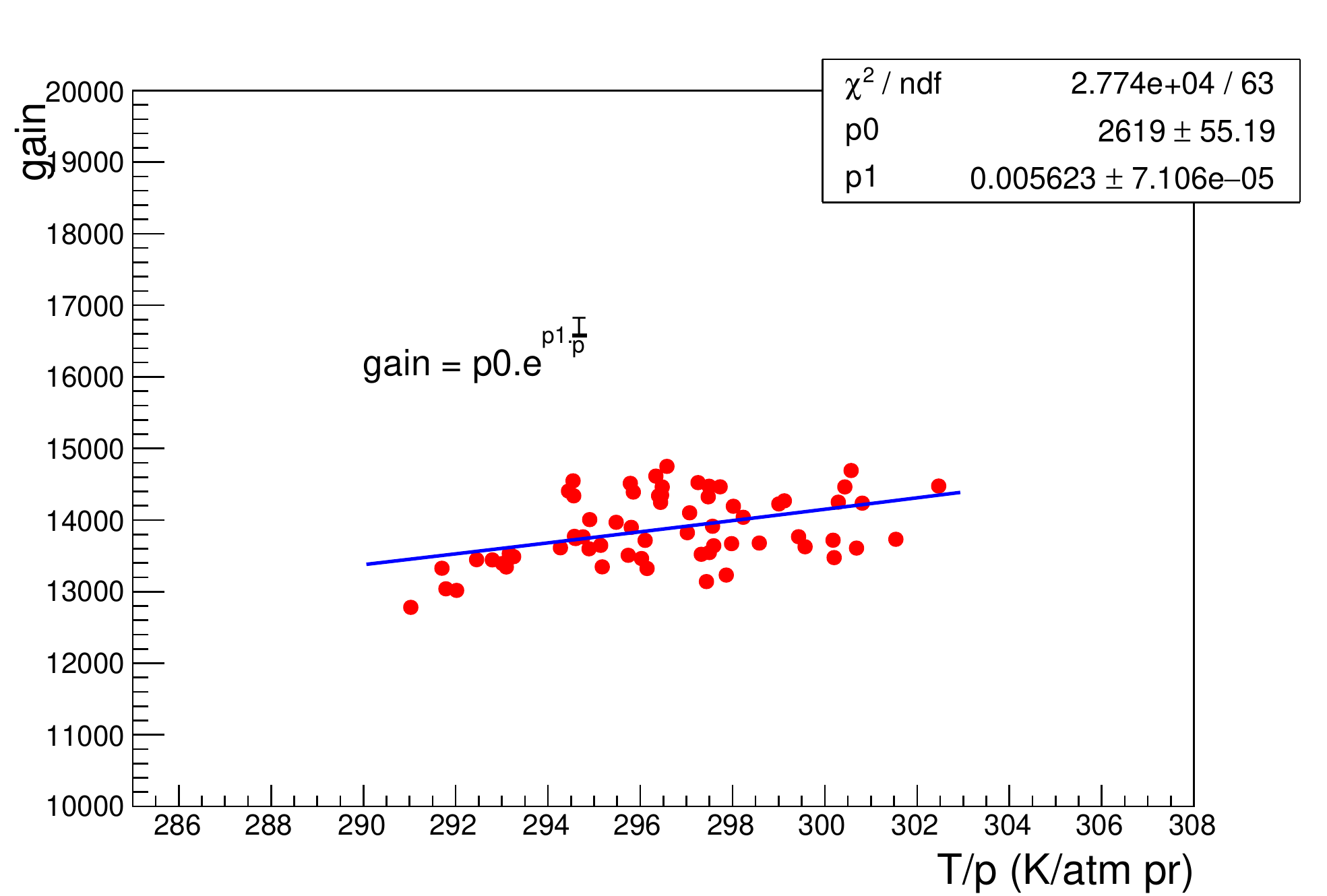}
\caption{Correlation between gain and T/p.}
\label{corr}
 \end{center}
\end{figure}

The gain of the straw tube is measured at regular intervals as mentioned earlier.
The gain as a function of time is shown in Fig.~\ref{gain} along with the variation of the ratio of ambient temperature (T=t+273~K) and pressure (p) with time. From Fig.~\ref{gain} it can be seen that during this period of more than 800~hrs, the gain decreased from 15000 to 13000. This may be the effect of prolonged radiation. 
 The aging rate is parameterized as a normalized gas gain loss:
 \begin{equation}\label{agerate}
     R = - \frac{1}{G_0} \frac{dG}{dQ} \times 100 \% \: per \: C/cm
 \end{equation}
where $G_0$ is the initial
gas gain, $dG$ is the loss of gas gain after collected charge $dQ$ per unit length.
To evaluate the aging rate, we normalised the instantaneous gain (gain$_{t}$) by the initial value of the gain (gain$_{t0}$) and plotted it against charge accumulated per unit length of the straw tube detector as shown in Fig.~\ref{normgain}.
The accumulated charge over the straw tube is calculated using the relation,
\begin{equation}
\frac{dQ}{dl} = \frac{r \times n \times e \times G \times dt}{dl}
\end{equation}
where $r$ is the measured rate (in Hz) incident on a particular area of the detector, $n$ is the
number of primary electrons for a single X-ray photon, $e$ is
the electronic charge, $G$ is the gain of the detector, $dt$ is the
time in second and $dl$ is the irradiated length of the straw. In this case the straw tube is exposed to continuous radiation for more than 800 hrs, owing to an accumulation of 0.6~C/cm of charge.
As shown in Fig.~\ref{normgain} the normalised gain is fitted by a 1$^{st}$ order polynomial function. Using the value of the slope (p1) the aging rate is calculated to be 9.6 $\%$ per C/cm. However, this observation needs a confirmatory test to ensure that this is purely due to radiation and no other external effects are responsible. It can be seen from Fig.~\ref{gain} that the variation of gain depends on variation in T/p. Although there is not much variation in T/p throughout the experiment, we still tried to find a correlation of gain with T/p and is shown in Fig.~\ref{corr}. It is seen from Fig.~\ref{corr} that the points are scattered and so the $\chi^2$/NDF of the fit is bad. Therefore, it can be said that there are other parameters also in addition to T/p, responsible for the variation of gain.
\begin{figure}[h!]
\begin{center}
\includegraphics[scale=0.35]{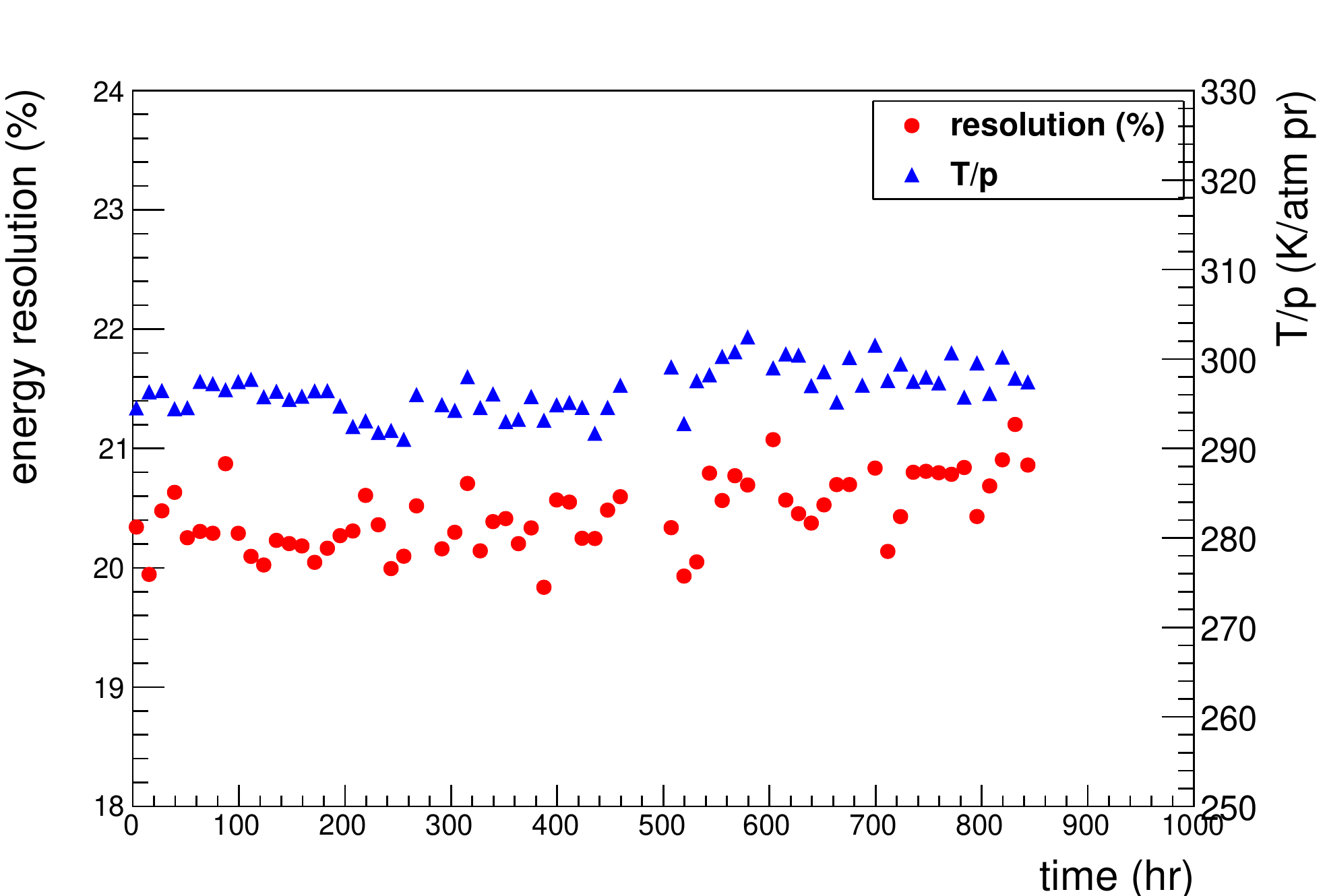}
\caption{ Energy resolution and T/p as a function of time. The bias voltage of the straw tube detector is 1550~V.}
\label{reso}
 \end{center}
\end{figure}

It is known that aging of gaseous detectors strongly depend on total accumulated charge \cite{kadyk2}. Apart from that, the aging rate is affected by macroscopic parameters such as, high gas gain, radiation intensity and gas flow rate. In that direction, the next experiment is carried out with high radiation intensity and low gas flow rates to observe aging rates for straw tube in a practically lesser amount of time.

Fig.~\ref{reso} shows the variation of the energy resolution with time. 
In this experiment, it is observed that the energy resolution increases from an initial value of 20$\%$ to a final value of 21$\%$ at the end of the measurement.


\section{Stability test of straws : Experiment II}
The goal of the second experiment was to verify whether the degradation in the gain of the straw tube detector is purely due to high irradiation or not. In order to verify this, gain and energy resolution measurements with two straws, marked as straw A and straw R are done. These two straws are positioned adjacent to one another. The idea is to use one straw as a reference detector (marked as R) and the other one as the detector under aging study (marked as A) such that there will be a much higher amount of charge accumulated on straw A than on straw R, after operating both the straws for a long duration. To study the performance, gain and energy resolution of both the straws are measured continuously and simultaneously at equal intervals of time. 
\subsection{Experimental set up}\label{expsetup2}
An identical experimental setup as mentioned in Section \ref{expsetup} is made to measure the gain and energy resolution of both the straws. The straws are connected to the same gas line such that any external factors affecting the performance of the straws cancel out when we take the ratio of any measured quantity of the two straws such as gain or energy resolution. The same Fe$^{55}$ X-ray source is used to irradiate both the straws. The radiation over straw R was purposely kept at low rates just to use it as a reference detector to monitor the gas gain continuously and compare at fixed time intervals with the gain of straw A. The particle rate from the source exposed to the straws is adjusted by using a perspex collimator. The count rates as measured by the scalar in case of straw A and straw R are 35~kHz/mm and 0.09~kHz/mm respectively. The biasing voltage of straw A and straw R are kept at 1550~V and 1450~V respectively. The reference straw, {\it i.e.}, straw R is operated at lower gains $\sim$6000 and lower radiation, whereas the straw under aging {\it i.e.}, straw A is operated at high gains $\sim$13000. This is done to achieve a higher amount of accumulated charge on straw A in comparatively lesser amount of time, whereas operating straw R at a low gain is done to minimise the amount of charge accumulation on it.
So the expectation is that the ratio of the gains of the two straws will normalise all the effects due to external parameters on the gas gain such as, ambient temperature, pressure, relative humidity and gas flow rate, except the effect of radiation which is different for both the straws. The energy spectra from straw A and straw R are simultaneously stored at regular intervals of time such that there is no time lag in between, with the help of two separate MCA modules. The detector characteristics such as gain and energy resolution are extracted from the Fe$^{55}$ X-ray spectra using the same method as discussed in Section~\ref{expsetup}.

\begin{figure}[h!]
\begin{center}
\includegraphics[scale=0.35]{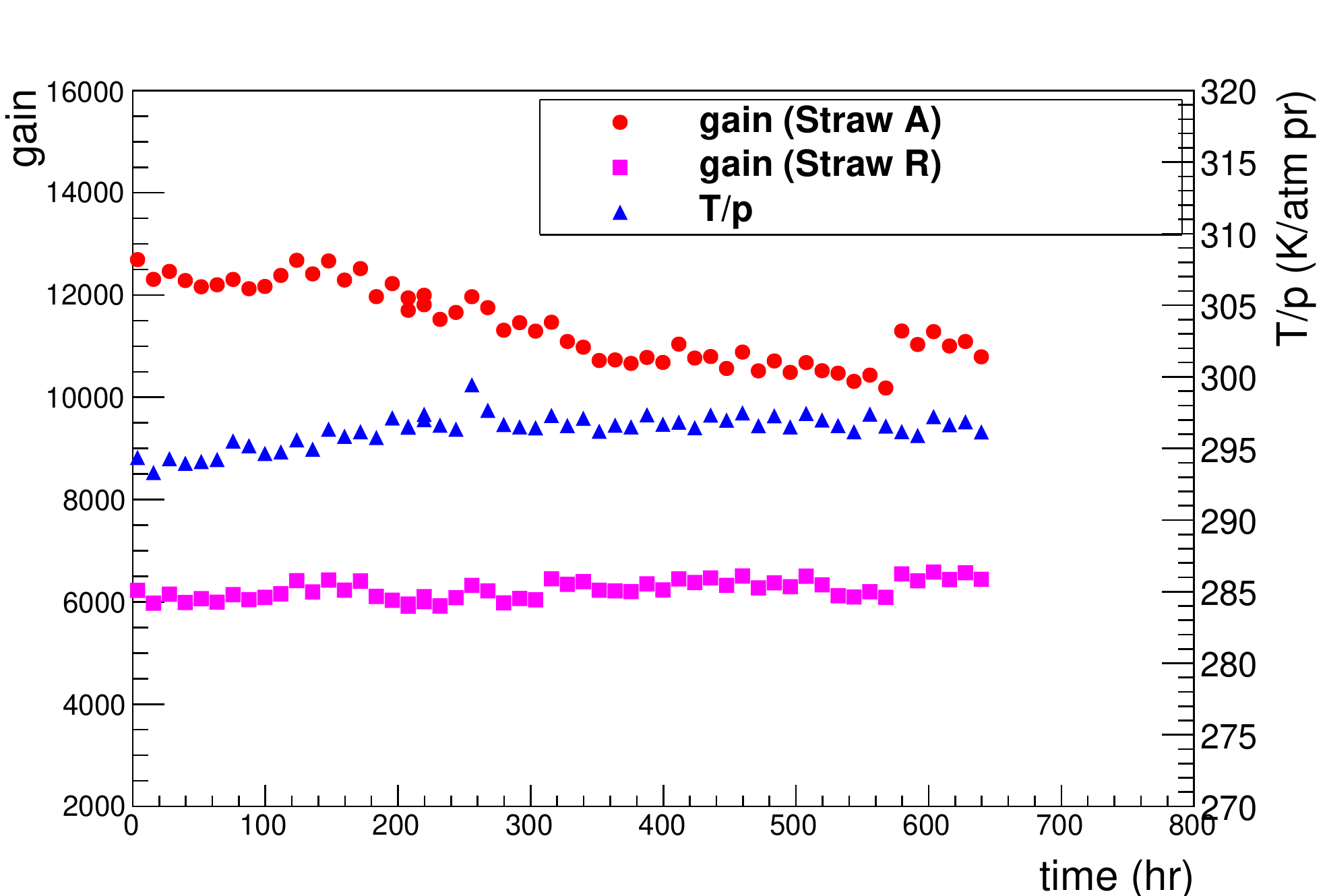}
\caption{Gain and T/p as a function of time for both the straws. The bias voltage of Straw A and R are 1550~V and 1450~V respectively.}
\label{gain56}
 \end{center}
\end{figure}

\begin{figure}[h!]
\begin{center}
\includegraphics[scale=0.35]{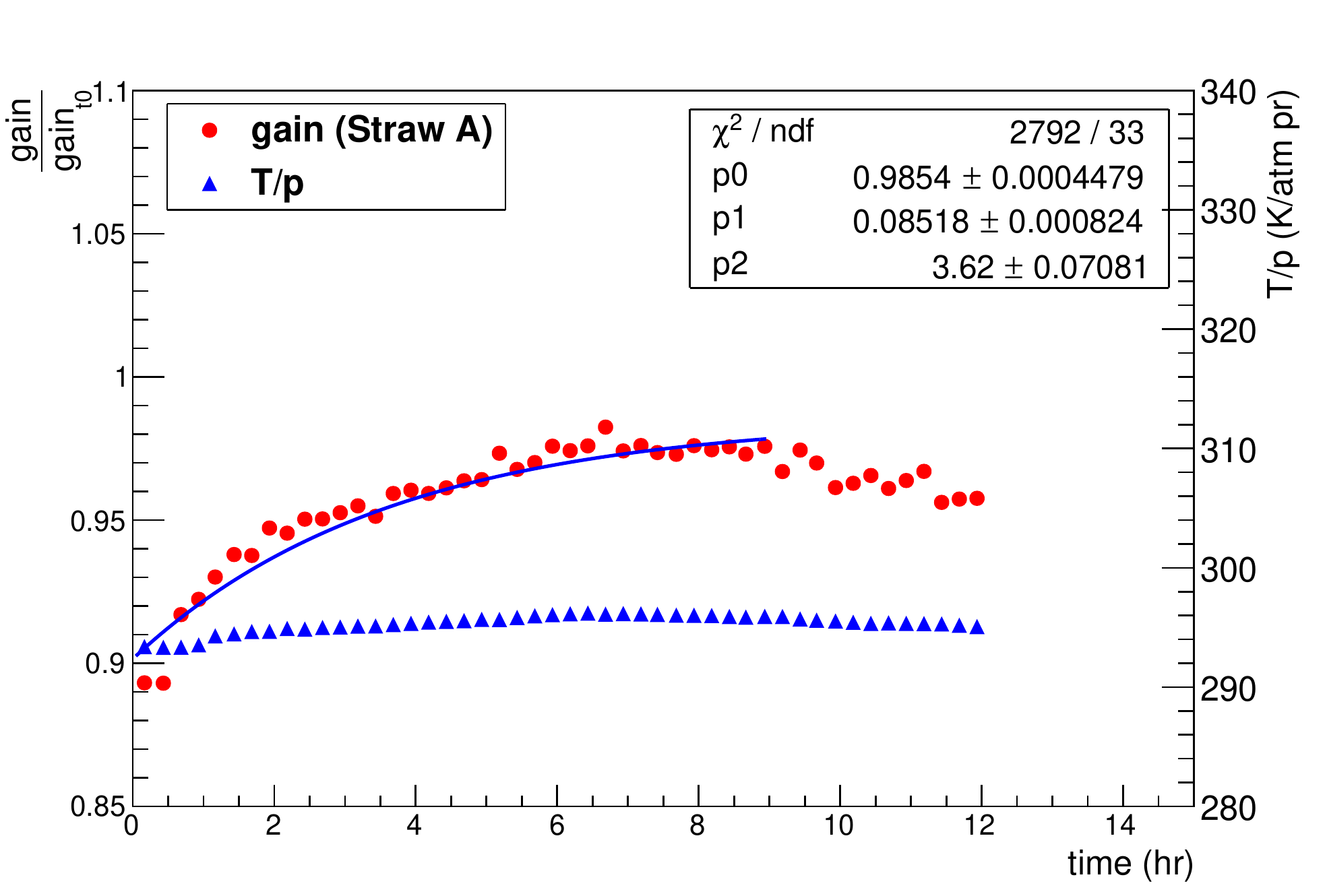}
\caption{Normalised gain and T/p as a function of time (here t=0 hr means t=100 hr of the actual measurement).}
\label{gainexpfit}
 \end{center}
\end{figure}

\subsection{Results}\label{res2}
\begin{figure}[h!]
\begin{center}
\includegraphics[scale=0.35]{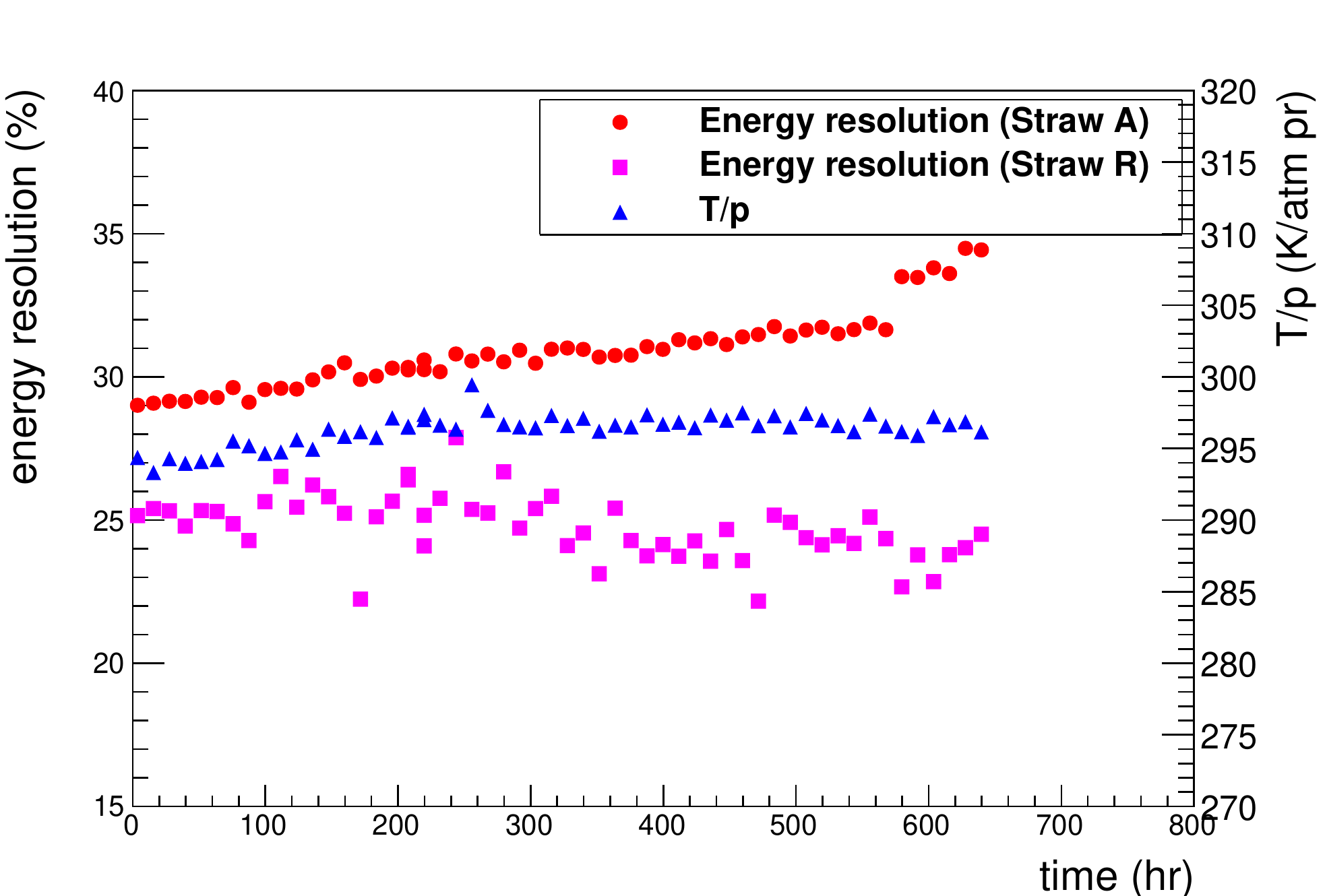}
\caption{Energy resolution and T/p as a function of time for both the straws. The bias voltage of straw A and R are 1550~V and 1450~V respectively.}
\label{reso56}
 \end{center}
\end{figure}

The variation of the gain and the energy resolution of both the straws over time is plotted in Fig.~\ref{gain56} and Fig.~\ref{reso56} respectively. Fig.~\ref{gain56} and Fig.~\ref{reso56} also show the variation of T/p that was also recorded throughout the experiment.
There is a gradual decrease in the gain of straw A with time as observed in Fig.~\ref{gain56}. However no such degradation in gain is observed for straw R. Whereas for straw A, in the first 100 hr of this measurement, we observed a decrease of 11~$\%$ in the absolute gain at gas flow rate of 0.02~l/hr. Then the gas flow rate was increased to $\sim$~0.13~l/hr at around time 100~hr, to check if the gain restores to its original value or not. In Fig.~\ref{gainexpfit}, the normalised gain of straw A, {\it viz.} the instantaneous gain (gain) over the initial gain (gain$_{t0}$ = 13000) and T/p is plotted as a function of time from the time instant the gas flow rate was increased. The time varying normalised gain is fitted by a function:
\begin{equation}\label{expfunction}
    f(t)=p_0(1-p_1e^{-\frac{t}{p_2}})
\end{equation}
where $p_0$, $p_1$ are constants, $t$ is the time in hr, and $p_2$ is the time constant of the function.

\begin{figure}[h!]
\begin{center}
\includegraphics[scale=0.35]{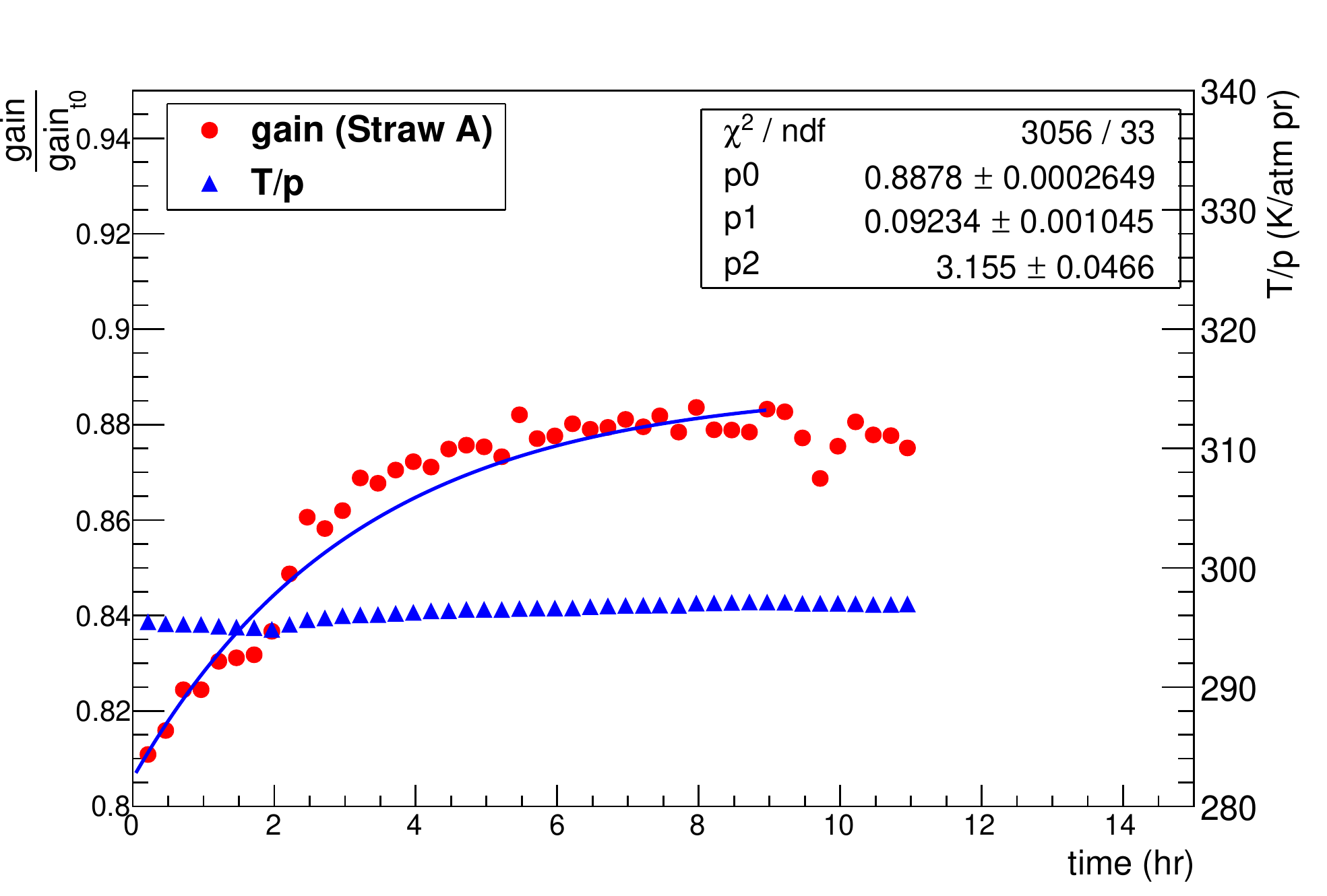}
\caption{Normalised gain and T/p as a function of time (here t=0 hr means t=600 hr of the actual measurement).}
\label{gainexpfit2}
 \end{center}
\end{figure}
From this fit it is seen that the gain restores to 96~$\%$ of its original value in 3.62~hrs of continuous gas flow at a rate of 0.13~l/hr. T/p was constant through out this time as can be seen from Fig.~\ref{gainexpfit} so the gain was not needed to be normalised by the T/p effect. After a few hours, the gas flow rate was again reset to a value $\sim$~0.03~l/hr and the measurement continued. The high voltage is kept ON and the source is not removed from its original position. We observed that the gain of straw A continuously decreased from 13000 to 10000 $viz.$ 80$\%$ of the original value of gain. We again increased the gas flow rate to a value of 0.8~l/hr at around time $\sim$ 600~hr, and observed an increase in the normalised gain as shown in Fig.~\ref{gainexpfit2}. The normalised gain is fitted with the same function as in eqn.~\ref{expfunction}. From the fit it is found that the gain increases from 80$\%$ to 87$\%$ of its original value in a time duration of 3.15~hrs, but the gain did not restore to its original value even after flowing the gas at a high rate for more than 10~hrs. The effect of changing the gas flow rate is also observed for straw R, but this effect is more prominent for the straw A. There is an overall slight increase in the gain of straw R which is due to the increase in T/p with time. For the same reason, there is a decrease in the energy resolution value of straw R from 25~$\%$ to 23~$\%$ as shown in Fig.~\ref{reso56}. However, it can be seen from Fig.~\ref{reso56} that the energy resolution of straw A increases from 29~$\%$ to 34~$\%$ (which is 17~$\%$ increment). It should also be noted here that the energy resolution of straw A does not improve on increasing the gas flow rate. This may indicate the performance degradation of the straw at high radiation environment. 


\begin{figure}[h!]
\begin{center}
\includegraphics[scale=0.35]{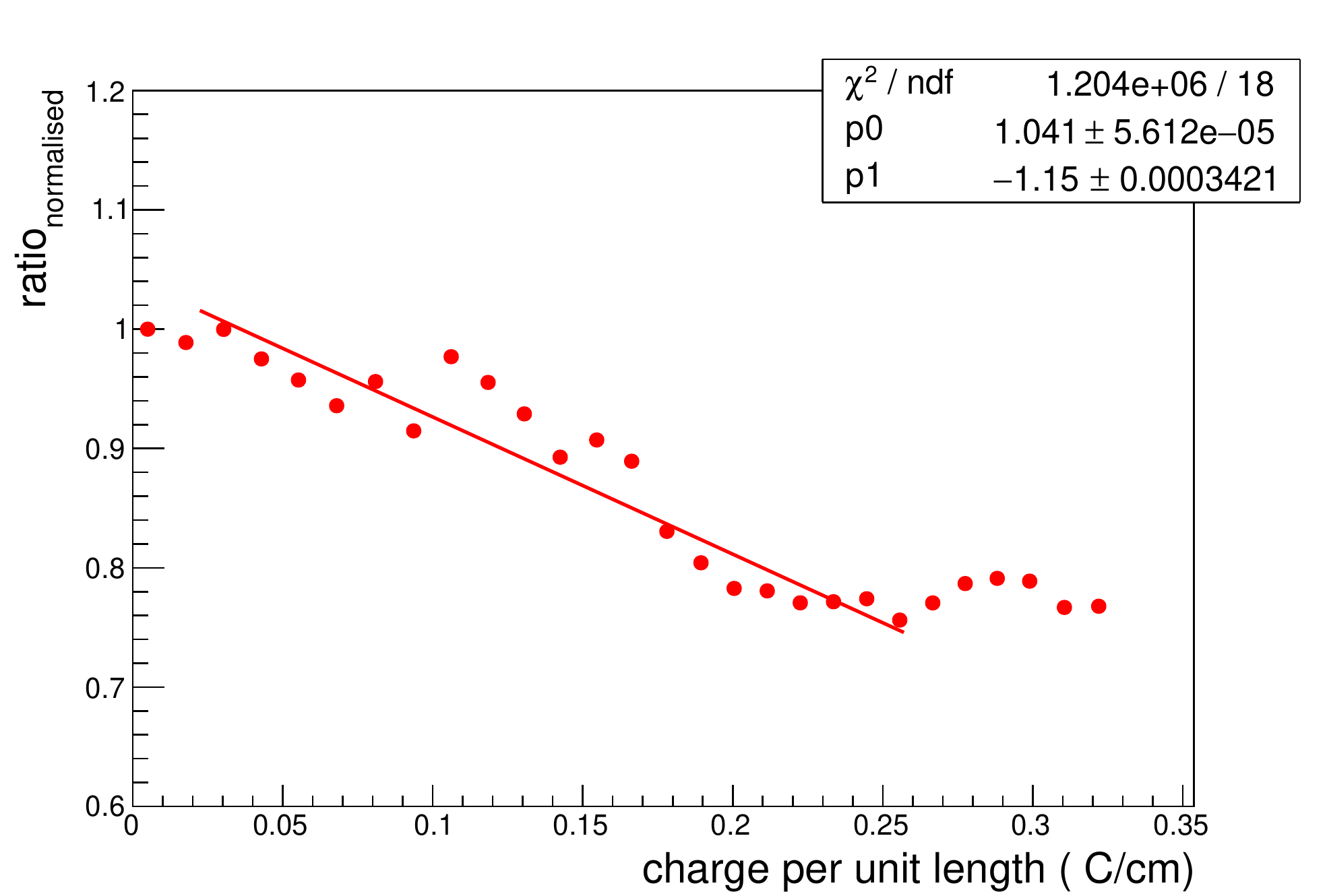}
\caption{Normalised ratio of gains of straw A and straw R as a function of charge accumulated per unit length of straw A.}
\label{ratio}
 \end{center}
\end{figure}

To understand the effect of high radiation alone on the degradation in the gain of the straw tube detector, we took the ratio of the gains of the straws and normalised it by the ratio of the initial gains. As changes in temperature, pressure, relative humidity and gas flow rate will affect both the straws similarly therefore, the ratio of the two gains can properly express the long term effect of the radiation only on the straw under aging study. 

The normalised ratio is calculated as 
\begin{equation}
ratio_{normalised} =\frac{\frac{gain_{straw A}(t)}{gain_{straw R}(t)}}{\frac{gain_{straw A}(0)}{gain_{straw R}(0)}}
\end{equation}
where $gain_{straw A}(0)$ and $gain_{straw R}(0)$ are the initial gains in straw A and straw R respectively.
 The normalised ratio is fitted by a 1$^{st}$ order polynomial function as shown in Fig.~\ref{ratio}.
 Here also a negative slope of -1.15 clearly indicates degradation of gain in straw A because of high rate of radiation.

\section{Measurement of Time resolution}\label{timemeas}
 Since straw tubes may be used for tracking in several upcoming high energy physics experiments so it is important to study its timing properties.
The time resolution of a gaseous detector depends on the gas mixture and applied voltage or electric field of the detector. It is actually the measure of the fluctuation in the time required for the electrons to drift along the electric field lines towards the anode wire. For wire chambers, the time resolution is usually of the order of a few {\it ns}.
\begin{figure}[h!]
\begin{center}
\includegraphics[scale=0.6]{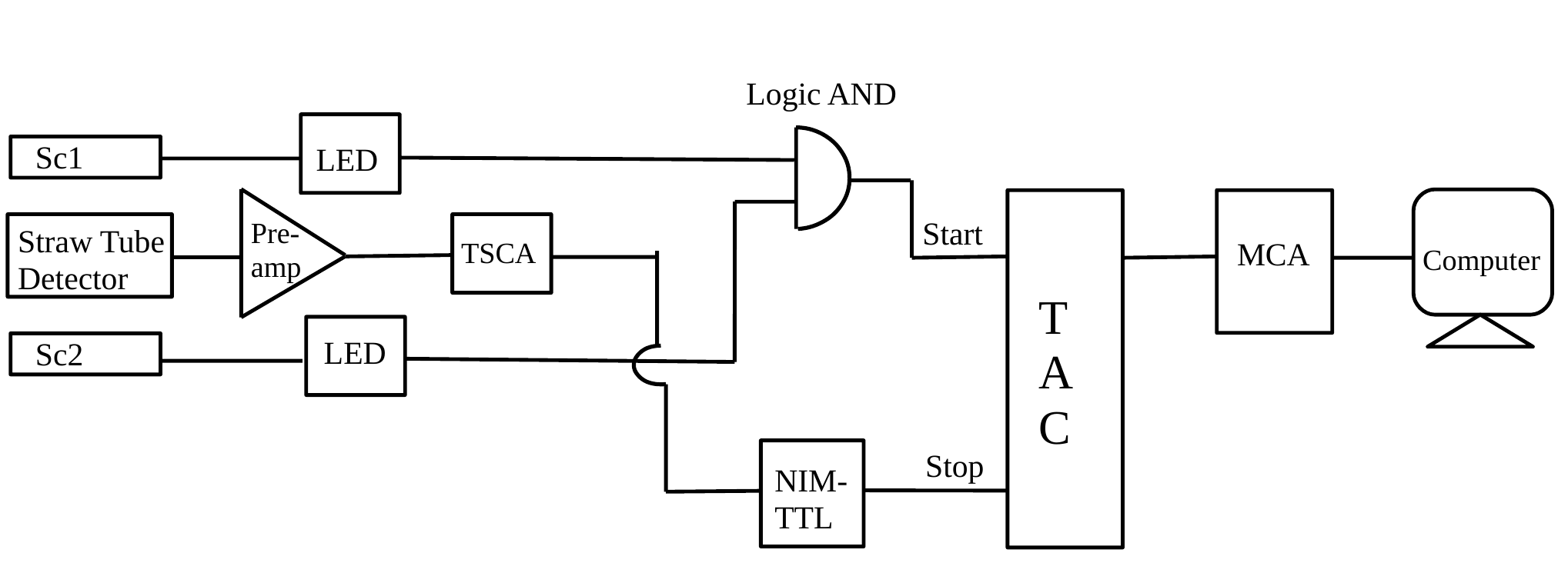}
\caption{Schematic of the electronics setup for time resolution measurement of the straw tube detector.}
\label{circuit}
 \end{center}
\end{figure}

The time resolution of the straw tube detector is measured with premixed Ar/CO$_2$ gas in the 70:30 volume ratio. Two plastic scintillator detectors are used to generate the 2-fold trigger signals with cosmic rays. The photomultipliers coupled to the scintillators are biased with a voltage of +~1550~V. The signals from the scintillators are fed to a Leading Edge Discriminator (LED) with a threshold of -~50~mV. The 2-fold coincidence signal is used as the `start' signal for the Time to Amplitude Convertor (TAC). The TAC is set at 10~$\mu$sec full scale range. The straw signal after passing through the preamplifier is fed to a Timing Single Channel Analyser (TSCA) which gives a TTL logic output. This TTL signal is converted to NIM signal using TTL-NIM adapter module and the NIM output signal is used as the `stop' signal for TAC.
\begin{figure}[h!]
\begin{center}
\includegraphics[scale=0.36]{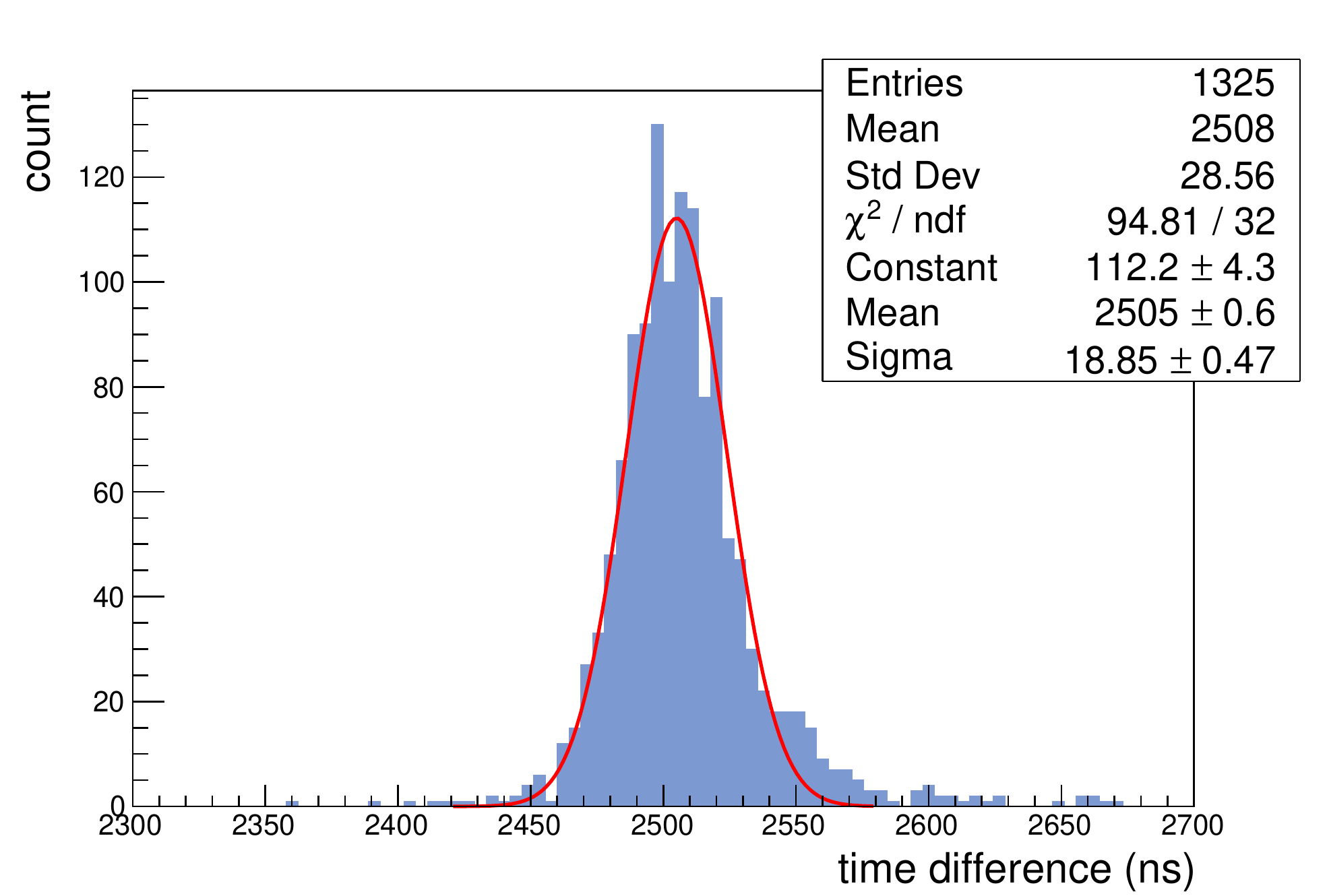}

\caption{Time spectrum of the straw tube detector at 1750~V with Ar/CO$_2$ 70:30 gas mixture. }
\label{timespec}
 \end{center}
\end{figure}
\begin{figure}[h!]
\begin{center}
\includegraphics[scale=0.36]{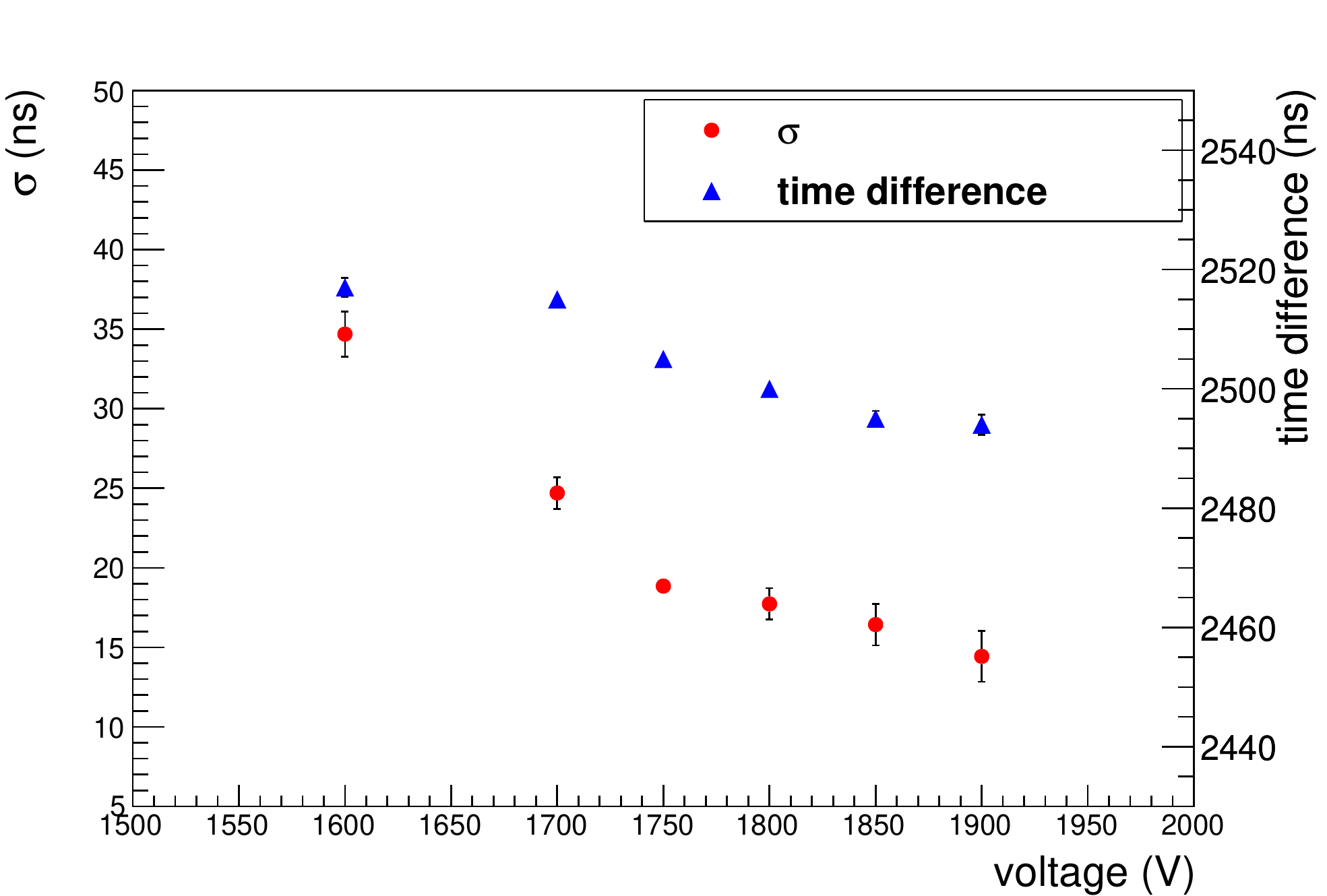}

\caption{Time resolution ($\sigma$) and time difference as a function of voltage.}
\label{timereso}
 \end{center}
\end{figure}
 The time difference between the `start' and the `stop' signal gets converted to amplitude in the TAC and the output is fed to the MCA for obtaining a timing spectrum. The schematic electronics setup for timing measurement is shown in Fig.~\ref{circuit}. Fig.~\ref{timespec} shows a typical time spectrum of a straw tube detector at 1750~V. The spectrum is fitted with a Gaussian function. The mean of the distribution gives the time difference of the trigger and the straw tube signal. The sigma of this distribution is the effective/combined time resolution of the straw tube and trigger detectors. Therefore using the relation:
\begin{equation}\label{eqntime}
\sigma_{eff}^2 = \sigma_{straw}^2 + \sigma_{1}^2 + \sigma_{2}^2 
\end{equation}
where $\sigma_{eff}^2$ is the effective time resolution of the combined detector setup, $\sigma_{1}$ and $\sigma_{2}$ are the time resolution of trigger scintillator Sc1 and scintillator Sc2 respectively, the time resolution of straw tube detector is extracted. The time resolution of the trigger scintillators were measured independently and the values of $\sigma_{1}$ and $\sigma_{2}$ were found out to be (0.38~$\pm$~0.01)~ns and (0.56~$\pm$~0.01)~ns respectively.
 The time resolution are measured for different voltage settings. The variation of the time resolution ($\sigma$) and the time difference as a function of voltage is shown in Fig.~\ref{timereso}. It is observed that the time resolution decreases with increasing voltage. The best achieved time resolution of straw tube detectors in our experiment using cosmic rays as the trigger is found to be (14.4~$\pm$~1.6)~ns at 1900~V.

\section{Summary and outlooks}\label{dis}
From the first experiment, we concluded that the gain reduction by 9.6$\%$ per C/cm after a total charge accumulation of 0.6~C/cm wire on the straw might be due to continuous and high radiation intensity. The observed aging rate is small but not negligible. In the second experiment, we confirmed that aging occurred due to the high radiation intensity and also drawn a conclusion on the dependence of this aging rate on the gas flow rate. It takes about 3~hrs time for the gain of a continuously irradiated straw to partially restore after increasing the gas flow rate. The transient nature of the aging is proven by the fact that the gain tends to restore as we increase the gas flow rate, but after a very long term exposure to radiation, we observe that the gain degrades continuously. The gain did not restore to its initial value even after flowing the gas at a very high rate through the straw chambers for a long time. This means that there is some aging due to long term operation of the straw tube detectors which was not observed in case of accelerated aging measurements reported in references~\cite{mcar,bond}. Therefore this needs further detailed investigation. For Ar/CO$_2$ gas mixture operated at high rates over long time periods, a gradual decomposition of CO$_2$ can occur and the resulting pure carbon can be deposited at the cathode \cite{age5}. An important observation in both the experiments is that the gain degradation of the straw tube detector starts immediately from the time of operation under high intensity radiation. The degradation is slow and gradual.
 Another conclusion that can be drawn from our experiment is that the straw tubes can be safely operated at low radiation intensities~($\sim$~0.1~kHz/mm) and at low gas flow rates~($\sim$~0.02~l/hr) (as no aging observed in straw R of experiment 2) and at high radiation intensities~($\sim$~40~kHz/mm) at high gas flow rates~($>$3~l/hr) (since slight gain reduction observed even after 800~hr of operation of the straws in experiment 1).
 A systematic study of the aging rates at different radiation intensities, gas gains and gas flow rates will be carried out in the future to fix the operating conditions of the straws in the real experiment.
 \\The time resolution of the straw tube detector is also measured with cosmic rays. The best achieved time resolution is found to be 14.4$\pm$1.6~ns at a biasing voltage of 1900~V.

\section{Acknowledgements}
We would like to thank Late Prof. Vladimir Peshekhonov of JINR, Dubna and Dr. Christian J. Schmidt
of GSI Detector Laboratory for providing the straw tube prototype.
We would like to thank Dr. S Chattopadhyay and Mr. J Saini of VECC, Prof. Rajarshi Ray, Prof. Somshubhro Bandyopadhyay and Dr. Sidharth K. Prasad of Bose Institute for valuable discussions and suggestions in the course of the study. This work is partially supported by the research grant SR/MF/PS-01/2014-BI from DST, Govt. of India and the research grant of CBM-MuCh project from BI-IFCC, DST, Govt. of India. S. Roy acknowledges her Institutional Fellowship research grant of Bose Institute. S. Biswas acknowledges the support of Intramural Research Grant provided by Bose Institute.

\end{document}